\documentclass[conference]{IEEEtran}
\IEEEoverridecommandlockouts
\newcommand{\herqules}{{\sc{herqules }}}
\newcommand{\subheading}[1]{\vspace{0.03in}{\noindent \textbf{#1.}}}

\usepackage{cite}
\usepackage{braket}
\usepackage{amsmath,amssymb,amsfonts}
\usepackage{algorithmic}
\usepackage{graphicx}
\usepackage{textcomp}
\usepackage{xcolor}
\usepackage{multirow}
\usepackage{makecell}
\usepackage[most]{tcolorbox}

\def\BibTeX{{\rm B\kern-.05em{\sc i\kern-.025em b}\kern-.08em
    T\kern-.1667em\lower.7ex\hbox{E}\kern-.125emX}}
\begin{document}

\title{Efficient and Scalable Architectures for Multi-level Superconducting Qubit Readout 
}

\author{

\IEEEauthorblockN{Chaithanya Naik Mude$^1$, Satvik Maurya$^1$, Benjamin Lienhard$^{2,3}$, Swamit Tannu$^1$}
\IEEEauthorblockA{$^1$Department of Computer Sciences, University of Wisconsin-Madison, Madison, WI 53706 USA}
\IEEEauthorblockA{$^2$Department of Chemistry, Princeton University, Princeton, NJ 08544 USA}
\IEEEauthorblockA{$^3$Department of Electrical and Computer Engineering, Princeton University, Princeton, NJ 08544 USA}
\vspace{-0.4in}
\thanks{B.L. is supported by the Swiss National Science Foundation (Postdoc.Mobility Fellowship grant \#P500PT\_211060).}
}

\maketitle

\begin{abstract}

Realizing the full potential of quantum computing requires large-scale quantum computers capable of running quantum error correction (QEC) to mitigate hardware errors and maintain quantum data coherence. While quantum computers operate within a two-level computational subspace, many processor modalities are inherently multi-level systems. This leads to occasional leakage into energy levels outside the computational subspace, complicating error detection and undermining QEC protocols. The problem is particularly severe in engineered qubit devices like superconducting transmons, a leading technology for fault-tolerant quantum computing. Addressing this challenge requires effective multi-level quantum system readout to identify and mitigate leakage errors. We propose a scalable, high-fidelity three-level readout that reduces FPGA resource usage by $60\times$ compared to the baseline while reducing readout time by 20\%, enabling faster leakage detection. By employing matched filters to detect relaxation and excitation error patterns and integrating a modular lightweight neural network to correct crosstalk errors, the protocol significantly reduces hardware complexity, achieving a $100\times$ reduction in neural network size. Our design supports efficient, real-time implementation on off-the-shelf FPGAs, delivering a 6.6\% relative improvement in readout accuracy over the baseline. This innovation enables faster leakage mitigation, enhances QEC reliability, and accelerates the path toward fault-tolerant quantum computing.

\end{abstract}

\section{Introduction}

Quantum computing offers the potential for significant computational speedups in fields like quantum chemistry, simulation, cryptography, and optimization, promising advantages over classical systems for tackling complex tasks. However, realizing these speedups depends on the efficient and scalable execution of quantum programs on robust hardware designed to support quantum operations. Quantum information is stored in inherently fragile qubits, the fundamental core units of quantum computation. These are highly susceptible to errors during gate operations due to device imperfections and environmental interference. This vulnerability to errors in qubit operations remains a fundamental challenge to advancing practical quantum technology.

Quantum Error Correction (QEC) can bridge the gap between error-prone qubit devices and practical quantum applications by encoding quantum information as logical qubits across multiple physical qubits to lower the overall error rate when the physical error rate is below a threshold. The effectiveness of QEC grows with redundancy, as measured by the code’s distance $(d)$, which exponentially suppresses errors, enabling QEC to achieve the low logical error rates required for practical quantum applications.

Superconducting qubit architectures are among the leading platforms for implementing QEC codes, such as scalable surface codes. QEC uses data qubits to store quantum information and parity qubits for measurements and parity checks, relying heavily on entangling gates and parity qubit measurements to detect and correct errors. A control system manages QEC operations by delivering precise gate pulses, leveraging FPGAs and signal generators for efficient operation in quantum systems with hundreds of qubits.

Readout is a fundamental operation in quantum computing. It is responsible for converting quantum information into classical information within the computational space, represented by the states ‘0’ and ‘1’. This readout process remains one of the most error-prone and slowest operation, highlighting the ongoing challenges in achieving practical, scalable superconducting quantum processors.

\begin{figure*}[t]
    \centering
    \includegraphics[width=0.8\linewidth]{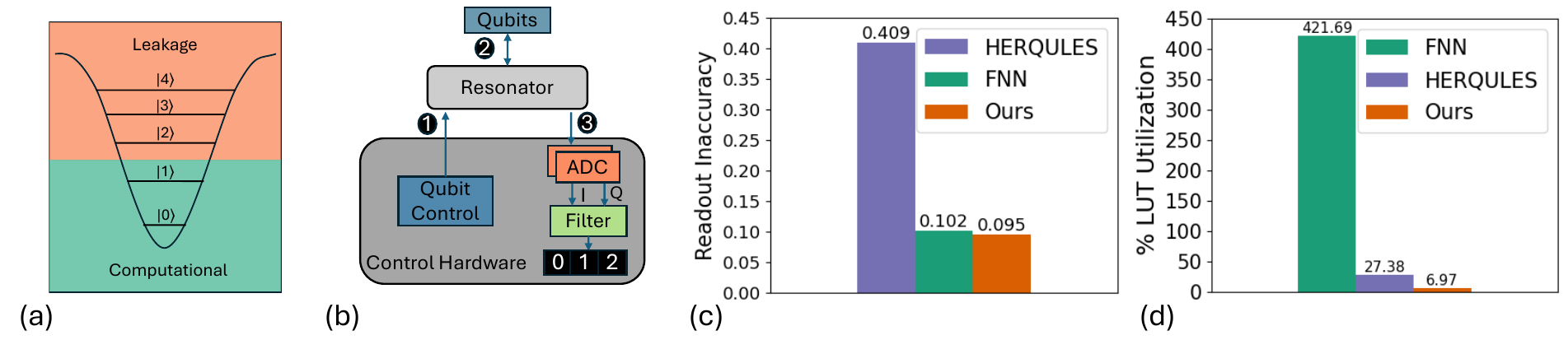}
    \vspace{-10pt}
    \caption{(a) Computational and Leakage levels in a qubit. (b) Overview of readout pipeline. (c) Comparison of readout classification inaccuracy over all five qubits used in Ref.~\cite{Blienhard}. (d) LUT utilization of using \herqules~\cite{HERCULES}, Feedforward Neural Network (FNN) design~\cite{Blienhard}, and our proposed method.}
    \label{fig:Intro}
    \vspace{-10pt}
\end{figure*}

Ideally, qubits in a quantum system should remain within their computational states, labeled ‘0’ and ‘1.’ However, due to the narrow energy gap between these computational states and higher energy levels, as shown in Fig.~\ref{fig:Intro}(a), qubits may transition to a higher, non-computational state, known as the leaked state ‘L.’ These leakage transitions, triggered by thermal excitations, quantum operations, or measurements, push qubits out of the computational basis. Leakage errors disrupt the function of quantum operations, often spreading to neighboring qubits. The effectiveness of QEC relies on precise and timely detection of errors through parity qubit measurements. Slow or inaccurate readout processes increasing the risk of leakage spreading across the system, jeopardizing QEC, potentially blocking the route to quantum advancement. Therefore, fast and effective detection and correction of leakage errors are essential for the reliability of QEC.

To mitigate leakage errors, specialized hardware elements called Leakage Reduction Circuits (LRCs)~\cite{leakage_supression_IBM, SC_leak_effect_QEC, DQLR, Markaida2020, Battistel_2021, LRB, McEwen_2021} are employed to restore qubits to the computational basis, thereby preserving qubit fidelity and maintaining QEC integrity. However, the effectiveness of LRCs hinges on the accuracy of leakage detection; if undetected, leakage can persist and lead to malfunctioning entangling gates employed in QEC circuits. Most LRCs depend on multi-level readout to reliably detect leakage states and apply corrective gates, while others require additional specialized control hardware~\cite{DQLR}. Multi-level readout also improves speculation of leakage on data qubits~\cite{ERASER} using the syndrome measurements of the surface code.




Beyond leakage mitigation, multi-level readout plays a critical role in expanding the capabilities of quantum systems. It enables qudit-based algorithms like efficient Toffoli decompositions~\cite{Litteken_2023, Nikolaeva_2023} and other complex computations. Despite improvements in qubit readout, reset~\cite{2023PhRvX..13d1057D, mesman2023qprofile}, reuse~\cite{brandhofer2023optimal}, and leakage errors continue to impact performance, highlighting the need for fast, scalable, and reliable multi-level readout to support error correction and advanced qudit algorithms.

Recent advancements in qubit-state readout accuracy have largely been driven by sophisticated discriminators, including deep neural networks (NN)\cite{Blienhard, azad2022machine}, hybrid approaches combining NNs with traditional methods\cite{HERCULES} such as matched filters~\cite{PhysRevA.91.022118}, and Hidden Markov Models~\cite{Varbanov_2020}. Feedforward NNs~\cite{Blienhard} and autoencoders~\cite{QutritReadout_autoencoders} can directly analyze digitized readout signals without pre-processing, capturing subtle data features that traditional methods often overlook. While these designs enhance state discrimination accuracy, their high computational demands limit scalability for multi-level readout.





While effective for two-level systems, existing solutions FNN~\cite{Blienhard}, \herqules\cite{HERCULES} often struggle to scale with multi-level readout due to large model architectures that face fidelity and hardware efficiency limitations, impacting leakage mitigation. LRCs are essential for addressing leakage errors, but imprecise applications can propagate faults. Fast, reliable leakage mitigation is crucial for fault-tolerant QEC. However, existing designs face two main limitations: large models that are too slow and scalable models~\cite{HERCULES} that quickly degrade in performance for multi-level systems. The Fig.~\ref{fig:Intro}(c), shows that \herqules is incapable of three-level readout. Additionally, Fig.~\ref{fig:Intro}(d) demonstrates that large models require significant FPGA resources, making implementation challenging. Our method uses fewer resources, performing better than the larger FNN model in readout discrimination accuracy, enabling efficient FPGA deployment.


This manuscript introduces a fast, scalable, and hardware-efficient three-level readout protocol. We reduce model size by almost $100\times$ over FNN~\cite{Blienhard} and $10\times$ over \textsc{herqules}~\cite{HERCULES}, providing quicker inference, enabling scalable, high-fidelity single-shot readout to advance the capabilities of multi-level qubit systems for rapid detection of leakage errors and improve reliability of QEC.

The key contributions of this paper are summarized below:
\begin{itemize}
    \item We propose a scalable multi-level readout protocol that uses a model size $100\times$ smaller and provides a 6.6\% relative improvement in accuracy over the baseline using matched filters and a modular lightweight neural network.
    \item Our design reduces hardware requirements significantly, utilizing $60\times$ fewer FPGA resources (LookUp Table (LUT)), thereby enabling efficient implementation on the off-the-shelf FPGA hardware.
    \item We enable a 20\% reduction in readout duration, enabling faster and more accurate leakage mitigation to improve overall system reliability.
\end{itemize}

\section{Background}




\subsection{Multi-Level Readout for Superconducting Qubits}



Multi-level readout is the process of determining a qubit’s state post-measurement, typically identifying it as the ground state ('0'), excited state ('1'), or leaked states ('L').  In superconducting qubits, this readout process is enabled by a dispersive coupling between qubits and resonators specifically used for qubit measurement~\cite{PhysRevA.69.062320, Yanagimoto_2023}. 

The readout pipeline, as shown in Fig.~\ref{fig:Intro}(b), consists of multiple stages: (1) the control hardware initiates a microwave probe tone sent to the resonator, (2) the qubit’s state induces a resonator phase shift picked up by the resonator probe tone, and (3) classical signal processing analyzes the transmitted or reflected readout resonator signal post frequency down-modulation and digitization to infer on the qubit state and assign a ground, excited, or leaked state label. This process is often slow and prone to errors, making precise state inference challenging. Achieving high superconducting-qubit-readout accuracies requires multiple analog components alongside robust signal processing. Here, we focus on enhancing the accuracy and scalability of qubit state discrimination. 


{\noindent \textbf{ADC.}} 
The incoming microwave signal is quadrature modulated, with its In-phase (I) and Quadrature (Q) components retrieved via analog mixing and digitized by two high-speed Analog-to-Digital Converters (ADCs) with typical sampling rates of 250-1000~MSamples/sec.

{\noindent \textbf{Filtering.}}
Due to the long measurement times, processing all time-bin samples generated by ADCs for classification is computationally and memory-intensive. Thus, most readout pipelines use a filtering scheme to condense this data. An averaging or a matched filter is commonly applied to reduce the I and Q data streams into a single representative value~\cite{PhysRevA.91.022118}.

{\noindent \textbf{Demultiplexing.}}
With frequency-multiplexed readout, qubits are divided into fixed groups to perform readout using the same physical channel. After filtering, the ADC samples are demultiplexed to determine the qubit's state within the group.

{\noindent \textbf{Classification.}}
The filtered, demultiplexed samples are used to classify the qubit's state using an appropriate classifier. 


\section{Impact of Qubit Leakage}
\subsection{Gate Malfunction due to Leakage}

Recent work by Google Quantum AI~\cite{DQLR, Acharya2023} shows that leakage errors are among the most significant error sources corrupting the logical qubit information. The impact is comparable to that of errors during Controlled-Z (CZ) gates. The error budget for leakage in CZ gates is similar to that of measurement and reset errors. Studies on superconducting architectures estimate the leakage probability to range from $10^{-4}$ to $10^{-3}$, suggesting that qubit leakage is relatively infrequent and random, making systematic investigations challenging.


We evaluate the effects of leakage experimentally using IBM quantum computers. We employ leakage injection techniques to assess leakage effects on the qubit gate performance, especially in Controlled-NOT (CNOT) gates widely used for surface code syndrome generation. Using the circuit of repeated CNOTs on the IBM Lagos, we initialize the control qubit in the leaked state ($\ket{2}$) and perform 10,000 shots with repeated CNOT operations to measure leakage instances in the target qubit. Results show significantly higher leakage growth of almost $3\times$ within 12 CNOTs with the leaked control qubit necessitating leakage removal.

In a single CNOT gate experiment with a leaked control qubit, with both $\ket{0}$ and $\ket{1}$ as target qubit states, we observed random bit flips and a leakage transfer of $1.5–2\%$ from the control qubit to target qubit after measuring the target qubit. 

\begin{tcolorbox}[colback=white!80!black,colframe=white!75!black]
   The presence of leakage \textit{malfunctions CNOT gates}, necessitating robust leakage mitigation strategies.
\end{tcolorbox}

\subsection{Impact on Leakage Speculation}

The characteristic bit-flip and leakage transport response of leaked qubits on CNOT gates can aid in speculating qubit leakage. Recent work, ERASER~\cite{ERASER}, utilizes surface code syndrome patterns to speculatively detect leakage and selectively apply LRCs, effectively reducing overall system leakage. Minimizing LRC usage is critical, as unnecessary applications can introduce additional leakage and non-leakage errors due to imperfect LRCs.


\begin{table}[ht]
\vspace{-10pt}
\normalsize
    \centering
    \caption{\normalsize Impact of readout on leakage speculation}
    \vspace{-5pt}
    \scalebox{0.9}{\begin{tabular}{|c|c|c|}
     \hline
         \textbf{Design} & \textbf{Accuracy}  & \textbf{Leakage Population}\\ \hline \hline
         ERASER & 0.957 &   4.19 $\times 10^{-3}$\\ \hline
         ERASER+M & 0.971 &   2.97 $\times 10^{-3}$\\ \hline
    \end{tabular}
    }
    \vspace{-5pt}
    \label{tab:speculation}
\end{table}

ERASER uses \textbf{Multi-level Readout} (ERASER+M) to capture leakage transport, improving leakage speculation accuracy by $2\%$ and leakage population (LP) by $1.5\times$ after 10 QEC cycles for a distance 7 surface code, as shown in Tab.~\ref{tab:speculation}.

\begin{tcolorbox}[colback=white!80!black,colframe=white!75!black]
  Multi-level readout enables leakage mitigation improving the performance of quantum error correction.
\end{tcolorbox}

\section{Scaling High-Fidelity Multi-Level Readout}
\label{sec:problem}

This section outlines the challenges in achieving scalable, high-fidelity single-shot readout for multi-level quantum systems and the limitations of existing readout methods.

\subsection{Factors affecting Single-Shot Readout Accuracy}
The single-shot readout fidelity captures the accuracy of determining a multi-level quantum state in a single measurement and is crucial for reliable quantum computations. Achieving high fidelity is essential for minimizing errors and enabling efficient quantum algorithms. Relaxation and excitation errors, due to crosstalk and unnecessary interactions, limit the fidelity.

\subheading{Relaxation Errors} 
Relaxation errors occur due to the spontaneous decay of higher-energy states during readout, caused by qubit-environment interactions. These errors are particularly problematic in long-latency readout operations. 


\subheading{Excitation Errors}
Excitation errors can occur when qubits are unintentionally excited to higher energy states during readout. The qubit in the ground state, $\ket{0}$, can get excited to $\ket{1}$ or higher. Similarly $\ket{1}$ can get excited to $\ket{2}$ or higher. 

\subheading{Crosstalk Errors} 
Readout crosstalk can occur in systems with multiple qubits and readout resonators in close spatial vicinity or frequency spacing. This effect causes the state of neighboring qubits to interfere with readout accuracy. Implementing a robust deep neural network demonstrated substantial error reduction by effectively mitigating the impact of crosstalk~\cite{Blienhard, QutritReadout_autoencoders}.

\subsection{Baseline Designs}
\begin{figure}
    \centering
    \includegraphics[width=\linewidth]{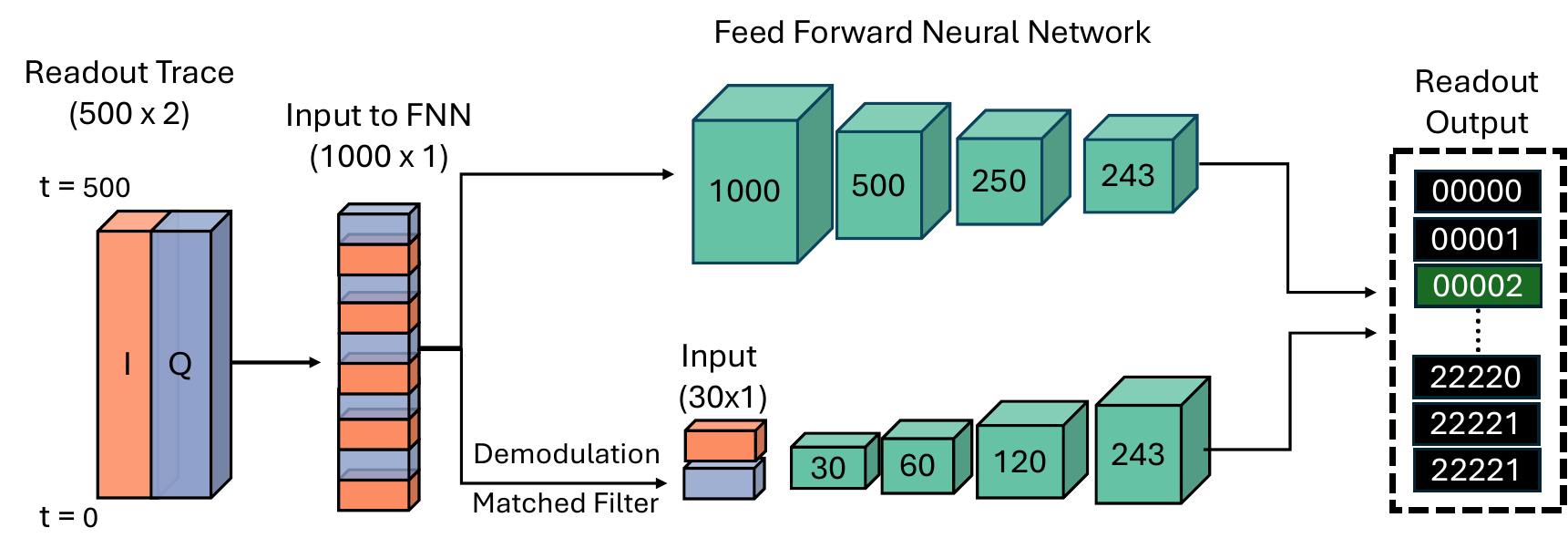}
    \caption{Design overview of FNN~\cite{Blienhard}(top) and {\sc herqules}~\cite{HERCULES}(bottom)  }
    \label{fig:baseline}
    \vspace{-10pt}
\end{figure}

\subheading{FNN Design~\cite{Blienhard}}
The intermediate-frequency readout signal is digitized and buffered before reaching a software classifier. Each readout trace, comprising 500 elements per I and Q channel sampled every $2 ns$ (totaling $1 \mu s$), serves as input to the FNN model~\cite{Blienhard}. To avoid undersampling, the model uses all ADC samples without demodulation, resulting in an input layer of 1000 neurons. The FNN has 32 outputs, representing the $2^{5}$ basis states of a five-qubit system. For our analysis with a 3-level quantum system, we modify the last layer to 243 outputs, representing $3^{5}$ basis states, as illustrated in Fig.~\ref{fig:baseline}.

\subheading{HERQULES Design~\cite{HERCULES}}
The ADC time-bin samples are demodulated to capture the patterns in traces corresponding to relaxation errors. After filtering and de-multiplexing, individual IQ values and assigned labels are used to obtain matched filters for relaxation and qubit states. This reduces the input size to $2\times$, the number of qubits with the output size as $2^{5}$. With a 3-level quantum system, the input layer increases to $6\times$ number of qubits and the output layer to 243 outputs to represent $3^{5}$ basis states, as illustrated in Fig.~\ref{fig:baseline} (bottom).


\subheading{Performance Analysis}
We examine the performance of state-of-the-art designs mentioned above for three-level quantum system readout. While the \herqules design outperforms FNN for two-level readout, it struggles with the increased complexity of three-level readout. In contrast, FNN achieves higher fidelity for three-level readout with a 686 thousand parameter model, but it cannot be efficiently implemented on an FPGA. Tab.~\ref{tab:baseline_performance} compares the readout fidelity of both designs, revealing performance degradation of \herqules in handling exponential increase in output states.


 \vspace{-5pt}
\begin{table}[ht]
\normalsize
    \centering
    \caption{\normalsize The three-level readout fidelity of existing state-of-the-art solutions with \textsc{\footnotesize $F_{5Q}$ = $\sqrt[\leftroot{-2}\uproot{2}5]{F_{1}F_{2}F_{3}F_{4}F_{5}}$}}
    \vspace{-5pt}
    \scalebox{0.75}{
    \begin{tabular}{|c|c|c|c|c|c|c|}
     \hline
         \textbf{Design} & \textbf{Qubit 1} & \textbf{Qubit 2}& \textbf{Qubit 3}& \textbf{Qubit 4}& \textbf{Qubit 5} & \textbf{$F_{5Q}$} \\ \hline \hline
         FNN &0.967   &0.728 &0.927 &0.932 &0.962& 0.898\\ \hline
         \herqules &0.598   &0.549 &0.608 &0.607&0.594& 0.591\\ \hline
         
    \end{tabular}
    }
    \label{tab:baseline_performance}
    \vspace{-8pt}
\end{table}
 
\subsection{Challenges with Existing Methods}

\subheading{Readout Accuracy} For multi-level systems, achieving high readout accuracy is challenging due to the complexity of distinguishing the exponentially large number of states compared to two-level systems. \herqules struggles with three-level systems due to a limited model capacity, and the FNN is impractical for real-time use due to its large parameter count and high hardware-demands highlighting the need for an accurate, hardware-efficient multi-level readout discriminator.

\subheading{Hardware Complexity}
The computational demands for implementing the FNN and \herqules designs grow with system size. For a system of $n$ qubits with $k$-levels each, the output layer scales exponentially as $k^{n}$ increasing total neural network parameters of the model. Additionally for \textsc{herqules}, the input layer scales as $O(nk^{2})$, growing quadratically with $k$ and linearly with $n$ due to the relaxation and error matched filters required between each pair of $k$-levels for each qubit.

\subheading{Readout Latency}
As quantum systems scale in the number of qubits or qudits, managing readout latency becomes critical for maintaining computational efficiency. Longer latencies can impair performance, particularly in large systems where timely feedback is essential for effective error correction and system stability. With increasing neural network parameters, inference latency also rises, limiting model designs for larger qubit counts. Since the output layer scales as $k^n$, models that scale linearly with the number of qubits $(n)$ are essential for practical implementation.

\subheading{Qubit Leakage Calibration}
Detection of leakage traces by calibrating them in a leaked state adds additional gate engineering steps and increases the resource overhead further.

\section{Enabling Efficient Multi-level Readout}

In this paper, we propose an architecture that tackles the challenges posed in the previous section and focuses on improving fidelity, reducing latency, and enhancing the scalability of superconducting quantum system readout.
\begin{figure}
    \centering
    \includegraphics[width=1\linewidth]{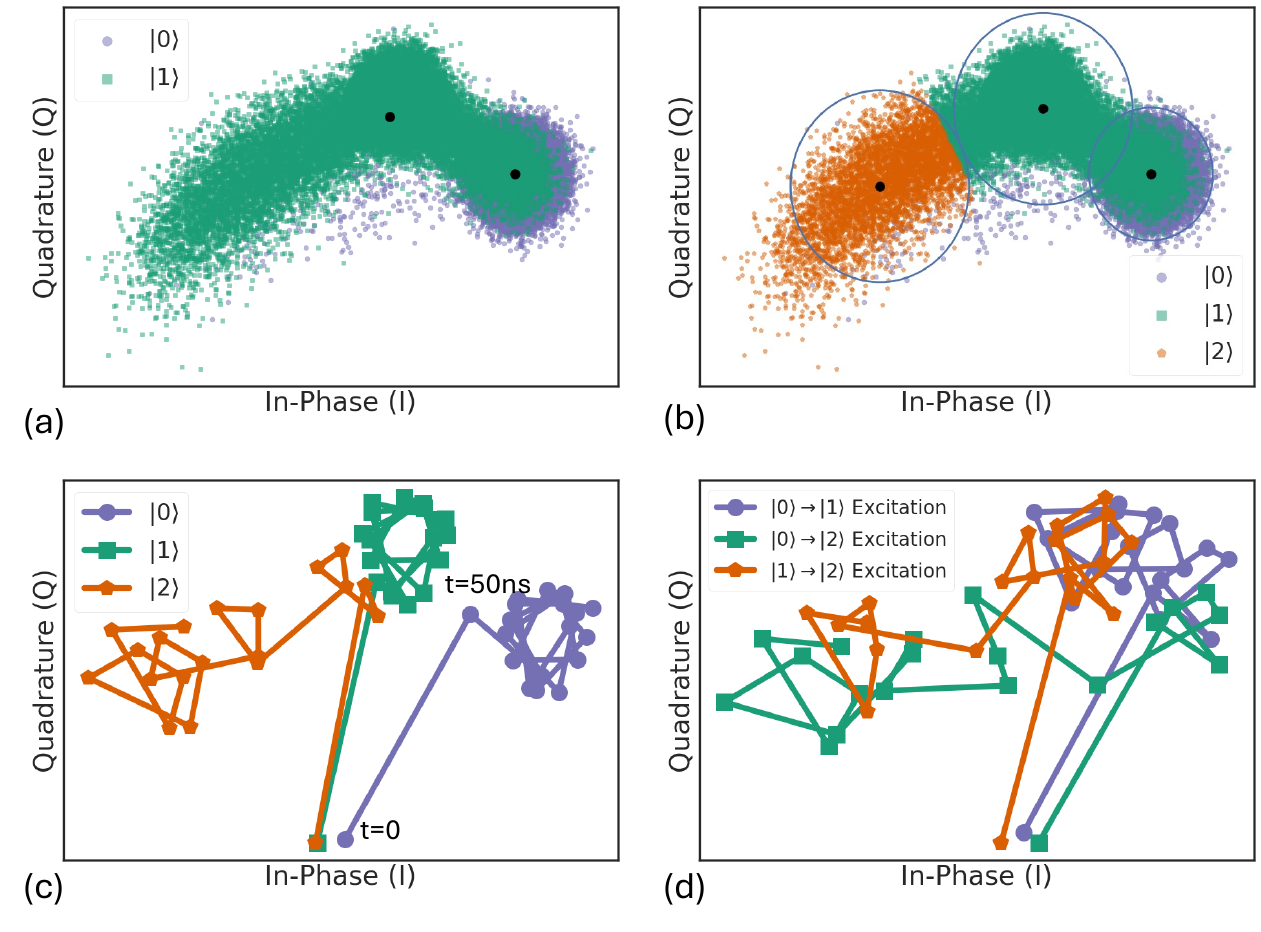}
    \vspace{-15pt}
    \caption{Averaged IQ data points for (a) two-level readout and (b) after detecting instances of natural leakage using spectral clustering. Mean traces of the clusters of (c) qubit states and (d) excitation error instances}
    \label{fig:traces_overview}
    \vspace{-15pt}
\end{figure}
\subsection{Detecting Leakage Cluster without Explicit Calibration}

\begin{figure*}[ht]
    \centering
    \includegraphics[width=\linewidth]{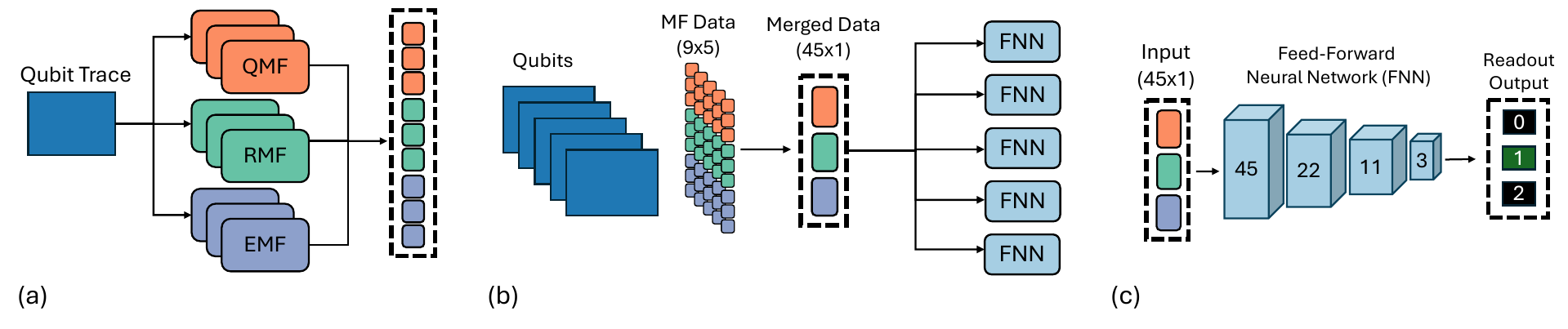}
    \vspace{-18pt}
    \caption{Overview of our Design. (a) The qubit traces are processed with matched filter envelopes by dot-product to output a single value for each envelope. (b) The data from all the qubits are merged to send the data as the input to a Feed-Forward Neural Network (FNN). (c) Our design of FNN to output qudit states.}
    \label{fig:our_design}
    \vspace{-10pt}
\end{figure*}
Calibrating qubits to populate leaked states is a complex process. Although rare, naturally occurring leakage in standard two-level readout traces reflects the probabilistic distribution of all leakage states more accurately than explicit calibration. These traces more accurately reflect the behavior of leaked qubits and can be distinguished from readout traces, as the averaged characteristics of readout traces for each state typically form distinct clusters.


We calculate the Mean Trace Value (MTV) to identify distinct clusters of qubit states as mentioned in \herqules~\cite{HERCULES}. For a trace $Tr$, MTV is defined as $MTV=\frac{1}{len(Tr)}\sum_{t=0}^{len(Tr)}Tr(t)$. This temporal mean of each trace corresponds to a single point in Fig.~\ref{fig:traces_overview}(a). Differences in the mean trace patterns across states suggest that readout trace-level information can enhance qubit state discrimination by leveraging inherent data patterns.

MTV points can be used to identify leaked states through spectral clustering into three classes. Most traces will correspond to computational states, while the smallest cluster will likely represent leaked states. Spectral clustering outputs three unlabeled clusters, which can then be labeled based on the probability of leakage when the state is prepared in a computational state, enabling accurate label assignment. As shown in Fig.~\ref{fig:traces_overview}(b), this approach identifies naturally occurring leakage states without needing explicit calibration.


\subsection{Matched Filter for Multi-Level Classification}


In single-qubit-state readout, matched filtering is a standard tool tomaximize the signal-to-noise ratio (SNR)~\cite{PhysRevA.91.022118}. Using statistical properties of signal traces, we define the Matched Filter (MF) kernel $K$ as the mean difference of traces normalized by variance differences, enhancing state discrimination by inversely weighting trace differences by variance. Let $\mu_0$ and $\mu_1$ represent the mean of traces corresponding to two distinct quantum states, with $\sigma_{0}^{2}$ and $\sigma_{1}^{2}$ as their respective variances. Then kernel is as $K=\frac{\mu_{1} - \mu_{0}}{\sigma_{1}^{2} - \sigma_{0}^{2}}$.

Most MFs are optimized for binary classification, relying on the statistical distinction between two states to maximize SNR. Higher-order MFs are mathematically very cumbersome. 
To address this, we use three two-state MF tailored to specific classes, but residual errors persist due to their limitations in separating multi-level states. To improve accuracy, we employ a small neural network to handle non-linearities.


\subheading{Deciphering Error Traces and Error Matched Filters}
The centroids of each state serve as priors for cluster identification. Traces belonging to a particular state but positioned closer to other cluster centroids can be tagged as error traces. With estimated centroids and state traces, we can use this information and ground truth data to label traces corresponding to relaxation and excitation events. Fig.~\ref{fig:traces_overview}(d) shows the MTV for such excitation traces from $\ket{0}$ to $\ket{1}$ and $\ket{2}$ and from $\ket{1}$ to $\ket{2}$. Quantum state discrimination can be improved by learning characteristic patterns exhibited by these error traces.


\subsection{Our Design}
The architecture of the FNN baseline design and \herqules learns features corresponding to crosstalk, qubit decay, and other non-idealities to achieve high-fidelity qubit readout. Unfortunately, they require significant computational and memory resources. We enable a hardware-efficient design that scales favourably with the increasing number of qubits.

The broad overview of the design is discussed in Fig.~\ref{fig:our_design}. Demodulated\footnote{demodulation is fast and in-expensive requiring two FMA units} ADC data is used to train MFs, including Qubit MF (QMF), Relaxation MF (RMF), and Excitation MF (EMF), as discussed in Tab.~\ref{tab:compare_mfs}. We incorporate a small 
NN to handle remaining non-linearities, yielding our design. The NN structure, shown in Fig.~\ref{fig:our_design}(c), has an input size (P) that scales as $O(nk^{2})$ for $n$ qubits with $k$-levels, as each qubit requires $O(k^{2})$ error MFs and the NN has two hidden layers with size $\lfloor P/2 \rfloor$, $\lfloor P/4 \rfloor$ and output size of $k$. For three-level systems, we have three QMFs, RMFs, and EMFs for each pair of levels summing up to an input size of 45 for our design.

\begin{table}[ht]
\normalsize
    \centering
    \caption{\normalsize Overview of employed Matched Filters}
    \vspace{-5pt}
    \scalebox{0.85}{\begin{tabular}{|c|c|}
     \hline
         \textbf{Matched Filter} & \textbf{To Distinguish}\\ \hline \hline
         Qubit Matched Filter (QMF) & $\ket{0}, \ket{1}, \ket{2}$\\ \hline
         Relaxation Matched Filter (RMF) & $\ket{1}\rightarrow \ket{0}, \ket{2}\rightarrow \ket{0}, \ket{2}\rightarrow \ket{1}$\\ \hline
         Excitation Matched Filter (EMF) & $\ket{0}\rightarrow \ket{1}, \ket{0}\rightarrow \ket{2}, \ket{1}\rightarrow \ket{2}$\\ \hline
         
    \end{tabular}
    \vspace{-15pt}
    }
    \label{tab:compare_mfs}
\end{table}

In contrast to \textsc{herqules}, which classifies all qubits collectively, our approach processes each qubit’s output individually while incorporating information from all qubits, resulting in $k$ outputs per qubit rather than $k^{n}$. This method enables polynomial growth in $(n,k)$ for model size rather than exponential, making it resource-friendly for larger qubit systems.

\subsection{Training Details}

MFs are based on the mean and variance of labeled readout traces with the aim to maximize the SNR for each state. We create qubit-level MFs for each pair of qubit states, as well as error MFs: RMFs for relaxation traces and EMFs for excitation traces. During inference, these kernels are applied to incoming readout traces (Fig.~\ref{fig:our_design}(a)) to generate likelihood scores for each state, which serve as inputs for further refinement. To address non-linearities, a NN is trained for each qubit on the outputs of qubit MFs, RMFs, and EMFs from all qubits using labeled data to optimize classification boundaries as shown in Fig.~\ref{fig:our_design}(b). During inference, the NN processes these outputs as shown in Fig.~\ref{fig:our_design}(c), refining state predictions, addressing overlaps, and enhancing multi-level readout accuracy.

\section{Methodology}

\subheading{Quantum Hardware}
We obtained datasets containing the readout time traces collected directly from the ADC originating from a five-qubit chip used in Ref.~\cite{Blienhard}. These qubits are read out via individual readout resonators coupled to a common feedline using frequency-multiplexing. The ADC sampling rate is 500 MSamples/sec, and qubit relaxation ($T_1$) times range from $7\mu s$ to $40\mu s$. 

The dataset contains readout traces for all 32 basis states of the five qubits, with 50,000 traces per basis state (32 × 50000 = 1600000 traces). We fixed the readout duration to $1\mu s$ for all qubits. Additionally, we use the third and fourth qubits, which are more prone to $\ket{2}$ excitations, to understand the impact of the excitation-matched filter.  

After spectral clustering, the total traces for computational and leaked states vary for each qubit, from the lowest of 487 traces for Qubit 1 to 17,642 for Qubit 4. We divide the train and test as 30-70 split for each of the $3^{5}$ possible states and use 15\% of the training dataset as the validation dataset. The distinguishability of the states of qubit 2 is limited due to the experimental setup in Ref.~\cite{Blienhard}.

\subheading{FPGA Hardware}
To estimate the FPGA resources needed to implement a NN, we use a combination of the \texttt{hls4ml}~\cite{fahim2021hls4ml} tool and Xilinx Vivado High-Level Synthesis (HLS). \texttt{hls4ml} can take a NN model written in frameworks such as Keras or Pytorch and create an equivalent HLS model that can then be synthesized with Vivado HLS. We use the Xilinx Zynq MPSoC \texttt{xczu7ev-ffvc1156-2-i} as the target device.

\section{Evaluations}


\subsection{Impact on Readout Fidelity}

Tab.~\ref{tab:compare_methods_all} presents the readout fidelity for the modified FNN design and our proposed method, showing a relative improvement of 6.6\% ($= \frac{90.52-89.85}{100-89.85}$). The FNN requires almost $85\times$ more LUTs than our method.
\begin{table}[ht]
\normalsize
    \centering
    \vspace{-5pt}
    \caption{\normalsize The three-level readout fidelity of all $3^{5}$ states with cumulative accuracy \textsc{\footnotesize $F_{5Q}$ = $\sqrt[\leftroot{-2}\uproot{2}5]{F_{1}F_{2}F_{3}F_{4}F_{5}}$}}
    \vspace{-5pt}
    \scalebox{0.85}{\begin{tabular}{|c|c|c|c|c|c|c|}
     \hline
         \textbf{Design} & \textbf{{\sc \small qubit 1}} & \textbf{{\sc \small qubit 2}}& \textbf{{\sc \small qubit 3}}& \textbf{{\sc \small qubit 4}}& \textbf{{\sc \small qubit 5}} & \textbf{$F_{5Q}$} \\ \hline \hline
         FNN &0.967   &0.728  &0.928 &0.932 &0.962 &0.8985 \\ \hline
         {\sc ours} &0.971 &0.745   &0.923 &0.939 &0.969&0.9052 \\ \hline
         
    \end{tabular}
    }
    \label{tab:compare_methods_all}
\end{table}


Qubit 3 and 4 are more prone to leakage, we want to compare our methods with existing single qubit methods. Tab.~\ref{tab:compare_methods} compares the readout fidelity of the discriminant-analysis based methods (LDA, QDA) and our proposed method. Our design demonstrates a $1-2\%$ improvement over NN and upto $6\%$ over LDA. This improvement is mainly attributed to additional information on relaxation and excitation errors.

         

\begin{table}[ht]
\normalsize
    \centering
    \caption{\normalsize The three-level readout fidelity of single-quantum states}
    \vspace{-10pt}
    \scalebox{0.85}{\begin{tabular}{|c|c|c|c|c|}
     \hline
         \textbf{Design} & \textbf{LDA} & \textbf{QDA} & \textbf{NN} & \textbf{OURS}\\ \hline \hline
         Qubit 3 & 0.8966 &0.914 &0.939 & 0.959\\ \hline
         Qubit 4 & 0.9181 &0.921 &0.926 &0.930\\ \hline
    \end{tabular}
    \vspace{-15pt}
    }
    \label{tab:compare_methods}
\end{table}



\subsection{Impact on Readout Latency and QEC Cycle Time}

We enable faster readout by reducing the readout time by 200 ns without much loss in the overall discrimination accuracy across all qubits at varying trace lengths, as shown in Fig.~\ref{fig:FPGA_fast_readout}b). Qubit-state readout, typically the slowest operation in QEC cycles, significantly reduces the QEC performance. The measurement-time reduction yields up to a 17\% decrease in QEC cycle time\footnote{QEC requires repeated measurements impacting execution time for quantum algorithms} for the surface-17 circuit~\cite{Versluis_2017}, providing a valuable tradeoff for systems with large code distances and directly decreasing total execution time.

\subsection{FPGA Resource Utilization}
\begin{figure}
    \centering
    \includegraphics[width=\linewidth]{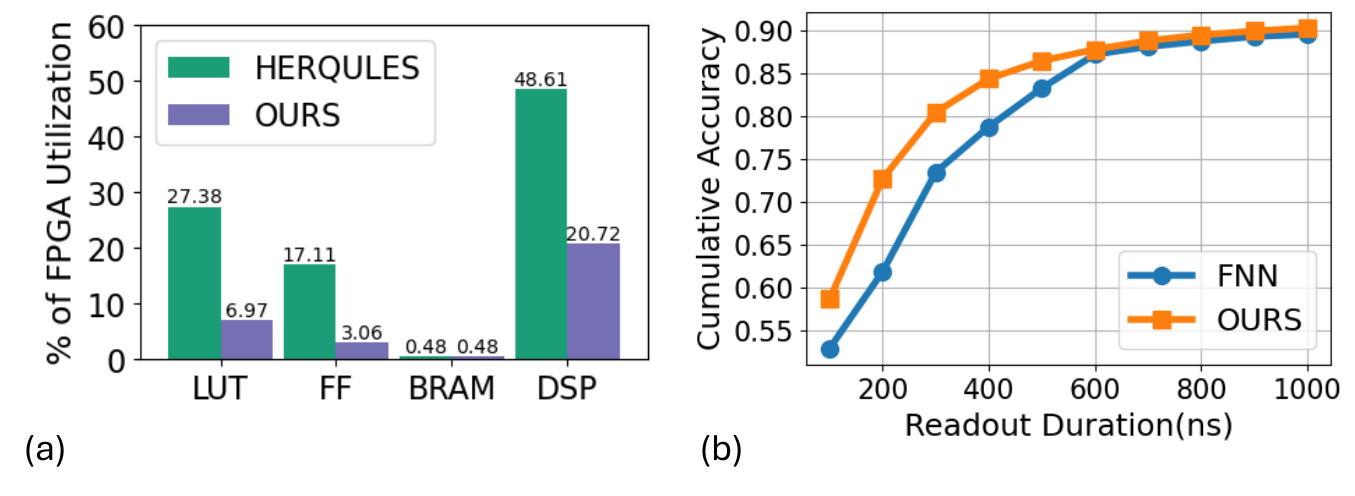}
    \vspace{-15pt}
    \caption{(a) Comparison of FPGA resource utilization (b)Variation of mean accuracy with readout duration in nanoseconds}
    \label{fig:FPGA_fast_readout}
    \vspace{-8pt}
\end{figure}

The FNN design requires $60\times$ $(\approx \frac{420}{7})$ more LUT utilization on an FPGA than our design and $15\times$ $(\approx \frac{420\%}{28\%})$ more than {\sc herqules}. Our design requires significantly lower FPGA resource utilization than {\sc herqules}, as illustrated in Fig.~\ref{fig:FPGA_fast_readout}(a), with the key metrics as LUTs, Flip-Flops (FF), Block RAM (BRAM), and Digital Signal Processing (DSP) units, with over $5\times$ reduction in FFs and $4\times$ LUTs compared to \herqules, indicating scalability of our approach. 

\subsection{Power Consumption}
The Synopsys design compiler is used to evaluated the power consumption using a 45nm TSMC standard cell library. With our design we require 1.561~mW total power at a 1~GHz clock rate and a latency of 5 cycles (5~ns).

\subsection{Impact on Leakage Speculation}

Our method and the FNN has fewer readout errors than discriminant-analysis-based methods, such as QDA and LDA. The accuracy of leakage speculation improves significantly as readout error decreases, rising from 0.913 to 0.947 as shown in Tab.~\ref{tab:leakage_speculation}. While the FNN outperforms QDA and LDA in speculation accuracy, it requires more inference time. Our method surpasses the FNN in both accuracy and speed, due to a $100\times$ smaller model size, enabling faster leakage detection and improving overall system performance. 

ERASER+M is run for 10 QEC cycles for a surface code to obtain speculation accuracy mentioned in Tab.~\ref{tab:leakage_speculation}. We calculate the error as the infidelity of mean accuracy excluding Qubit 2 due to experimental limitations during its setup.

\begin{table}[hpt]
\begin{center}
\begin{small}
\vspace{-0.15in}
\caption{Impact of multi-level readout on leakage speculation}
\label{tab:leakage_speculation}
\setlength{\tabcolsep}{0.05cm} 
\renewcommand{\arraystretch}{1.2}
\vspace{-0.1in}
\scalebox{0.85}{
\begin{tabular}{|c|c|c|c|}
\hline
\centering
\textbf{\quad Design \quad }& \textbf{Error(\%)}  & \textbf{\quad Speed \quad}     & \textbf{Speculation Accuracy} \\ \hline \hline
  LDA & \textbf{10} & Fast          & 0.914        \\ \hline 
QDA & \textbf{9} & Fast           & 0.921      \\ \hline 
FNN  & \textbf{5.5} & Slow       & 0.943   \\ \hline
Ours & \textbf{5} & Fast      & 0.947   \\ \hline
\end{tabular}
}
\vspace{-0.15in}
\end{small}
\end{center}
\end{table}

\section{Conclusion}

We present a scalable, hardware-efficient qudit-state-readout protocol that combines matched filters with lightweight neural networks to achieve high accuracy and efficient leakage mitigation. By transitioning the scaling of the neural network architecture from exponential to polynomial, our approach reduces hardware demands and enables practical FPGA deployment. Additionally, enabling fast readout with a 20\% reduction in readout duration accelerates performance without requiring additional training. This multi-level readout design strengthens QEC by enabling effective speculation for fast leakage detection, advancing reliable, fault-tolerant quantum systems and moving closer to efficient quantum processors.


\bibliography{References}
\bibliographystyle{ieeetr}

\end{document}